\documentclass[acus]{JAC2000}
\usepackage{graphicx}
\begin{document}

\title{A SECOND-ORDER STOCHASTIC LEAP-FROG ALGORITHM 
FOR LANGEVIN SIMULATION\thanks{Work supported by
DOE Grand Challenge in Computational Accelerator Physics,
Advanced Computing for 21st Century Accelerator Science and Technology
Project, and Los Alamos Accelerator Code Group using resources at
the Advanced Computing Laboratory and the National Energy
Research Scientific Computing Center.}
}

\author{Ji Qiang and Salman Habib, LANL,
         Los Alamos, NM 87545, USA}

\maketitle

\begin{abstract}
Langevin simulation provides an effective way to study collisional
effects in beams by reducing the six-dimensional Fokker-Planck
equation to a group of stochastic ordinary differential equations.
These resulting equations usually have multiplicative noise since the
diffusion coefficients in these equations are functions of position
and time.  Conventional algorithms, e.g. Euler and Heun, give only
first order convergence of moments in a finite time interval.
In this paper, a stochastic leap-frog algorithm for the numerical
integration of Langevin stochastic differential equations with
multiplicative noise is proposed and tested. The algorithm has a
second-order convergence of moments in a finite time interval and
requires the sampling of only one uniformly distributed random variable
per time step. As an example,
we apply the new algorithm to the study of a mechanical oscillator
with multiplicative noise.
\end{abstract}

\section{INTRODUCTION}
Multiple Coulomb scattering
of charged particles, also called intra-beam scattering, has important
applications in accelerator operation. It causes a diffusion
process of particles and leads to an increase of beam size and emittance. 
This results in a fast decay of the quality of beam and
reduces the beam lifetime when the size of the beam is large enough to
hit the aperture~\cite{piwinski}.

An appropriate way to study the multiple Coulomb scattering is to
solve the Fokker-Planck equations for the distribution function in
six-dimensional phase space. 
Nevertheless, the Fokker-Planck equations 
are very expensive to solve numerically
even for dynamical systems possessing only a very modest number of
degrees of freedom. Truncation schemes or closures 
have had some success in extracting the behavior of
low-order moments, but the systematics of these approximations remains
to be elucidated. On the other hand, the Fokker-Planck equations
can be solved using an equivalent Langevin simulation, which reduces the
six-dimensional partial differential equations into a group of stochastic
ordinary differential equations. Compared to the Fokker-Planck equation, 
stochastic differential equations are not difficult to solve, 
and with the advent
of modern supercomputers, it is possible to run very large numbers of
realizations in order to compute low-order moments accurately.
In general, the noise in these stochastic ordinary differential equations 
are multiplicative instead of additive since
the dynamic friction coefficient and diffusion coefficient
in the Fokker-Planck equations depend on the spatial position.
An effective numerical algorithm to integrate the stochastic differential
equation with multiplicative noise will significantly improve the 
efficiency of large scale Langevin simulation.

The stochastic
leap-frog algorithms in the Langevin simulation 
are given in Section~II. Numerical
tests of this algorithms 
is presented in Section~III. A physical application of the algorithm
to the multiplicative-noise mechanic oscillator is given in
Section~IV. The conclusions are drawn in Section~V.

\section{STOCHASTIC LEAP-FROG ALGORITHM}
In the Langevin simulation, 
the stochastic particle equations of motion
that follow from the Fokker-Planck equation are (Cf. Ref. \cite{risken})
\begin{eqnarray}
{\bf r}' & = & {\bf v}, \\
{\bf v}' & = & \frac{{\bf F}}{m} - \nu {\bf v} + \sqrt{D}
{\bf \Gamma}(t), \label{langevin}
\end{eqnarray}
where ${\bf F}$ is the force 
including both the external force and the self-generated mean field
space charge force, $m$ is the mass of particle, $\nu$ is friction coefficient,
$D$ is the diffusion coefficient, 
and
${\bf \Gamma}(t)$ are Gaussian random variables with
\begin{eqnarray}
\langle \Gamma_i(t)\rangle & = & 0,  \\
\label{noise1}
\langle \Gamma_i(t)\Gamma_i(t') \rangle & = & \delta(t-t').
\label{noise2}
\end{eqnarray}
In the case not too far from thermodynamic equilibrium,
the friction coefficient is given as
\begin{equation}
\nu = \frac{4\sqrt{\pi}n({\bf r})Z^4e^4 \ln{(\Lambda)}}{3 m^2(T({\bf r})/m)^{3/2}}
\end{equation}
and the diffusion coefficient $D$ is $D=\nu kT/m$~\cite{jones}. 
Here, $n({\bf r})$ is the density of particle, $T({\bf r})$ 
is the temperature of
of beam, $Z$ is the charge number of particle, $e$ is the charge of
electron, $\Lambda$ is the Coulomb logarithm, and $k$ is the Boltzmann
constant.
For the above case, noise terms enter only in the
dynamical equations for the particle momenta. In Eqn.~(\ref{bmeqs})
below, the indices are single-particle phase-space coordinate indices;
the convention used here is that the odd indices correspond to
momenta, and the even indices to the spatial coordinate. In the case
of three dimensions, the dynamical equations then take the general
form:
\begin{eqnarray}
\dot{x}_1 & = & F_1(x_1,x_2,x_3,x_4,x_5,x_6) +
\sigma_{11}(x_2,x_4,x_6) \xi_1(t) \nonumber \\
\dot{x}_2 & = & F_2(x_1) \nonumber \\
\dot{x}_3 & = & F_3(x_1,x_2,x_3,x_4,x_5,x_6) +
\sigma_{33}(x_2,x_4,x_6) \xi_3(t) \nonumber \\
\dot{x}_4 & = & F_4(x_3) \nonumber \\
\dot{x}_5 & = & F_5(x_1,x_2,x_3,x_4,x_5,x_6) +
\sigma_{55}(x_2,x_4,x_6) \xi_5(t) \nonumber \\
\dot{x}_6 & = & F_6(x_5)
\label{bmeqs}
\end{eqnarray}
In the dynamical equations for the momenta, the first term on the
right hand side is a systematic drift term which includes the effects
due to external forces and damping. The second term is stochastic in
nature and describes a noise force which, in general, is a function of
position. The noise $\xi(t)$ is first assumed to be Gaussian and white
as defined by Eqns.~(3)-(\ref{noise2}). The stochastic
leap-frog algorithm for Eqns.~(\ref{bmeqs}) is written as
\begin{eqnarray}
\bar{x}_i(h) & = & \bar{D}_i(h) + \bar{S}_i(h)
\end{eqnarray}
The deterministic contribution $\bar{D}_i(h)$ can be obtained using
the deterministic leap-frog algorithm.
Here, the
deterministic contribution $\bar{D}_i(h)$ and the stochastic
contribution $\bar{S}_i(h)$ of the above recursion formula for
one-step integration are found to be
\begin{eqnarray}
\bar{D}_i(h) & = & \bar{x}_i(0) +
h F_i(\bar{x}_1^*,\bar{x}_2^*,\bar{x}_3^*,\bar{x}_4^*,\bar{x}_5^*,
\bar{x}_6^*);  \mbox{ \hspace{0.5cm}} \nonumber\\
&& \{i=1,3,5\} \nonumber \\
\bar{D}_i(h) & = & \bar{x}_i^* \nonumber\\
&& + \frac{1}{2} h
F_i\left[x_{i-1}+hF_{i-1}(\bar{x}_1^*,\bar{x}_2^*,\bar{x}_3^*,
\bar{x}_4^*,\bar{x}_5^*,
\bar{x}_6^*)\right];\nonumber\\
&& \{i=2,4,6\} \nonumber \\
\bar{S}_i(h) & = &  \sigma_{ii}\sqrt{h} W_i(h) +
\frac{1}{2}F_{i,k}\sigma_{kk} h^{3/2} \tilde{W}_i(h)\nonumber\\
&& +\frac{1}{2} \sigma_{ii,j}F_j h^{3/2} \tilde{W}_i(h) \nonumber \\
& & +\frac{1}{4}F_{i,kl}\sigma_{kk}\sigma_{ll} h^2 \tilde{W}_i(h)
\tilde{W}_i(h); \nonumber\\
&& \{i=1,3,5;~ j=2,4,6;~ k,l = 1,3,5\}
\nonumber \\
\bar{S}_i(h) & = &
\frac{1}{\sqrt{3}}F_{i,j}\sigma_{jj}h^{3/2}\tilde{W}_j(h) \nonumber\\
&&+\frac{1}{4}F_{i,jj}\sigma_{jj}^2
h^2\tilde{W}_j(h)\tilde{W}_j(h)\nonumber\\
&&\{i=2,4,6;~ j=1,3,5\}  \nonumber \\
\bar{x}_i^* & = & \bar{x}_i(0) + \frac{1}{2} h F_i(\bar{x}_1,\bar{x}_2,
\bar{x}_3,\bar{x}_4,\bar{x}_5,\bar{x}_6)\nonumber\\
&& \{i = 1,2,3,4,5,6\}
\label{walgo}
\end{eqnarray}
where $\tilde{W}_i(h)$ is a series of random numbers with the moments
\begin{eqnarray}
\langle\tilde{W}_i(h)\rangle &=&\langle(\tilde{W}_i(h))^3\rangle =
\langle(\tilde{W}_i(h))^5\rangle = 0  \\
\langle(\tilde{W}_i(h))^2 \rangle & = & 1,~~~\langle(\tilde{W}_i(h))^4
\rangle = 3
\end{eqnarray}
This can not only be achieved by choosing true Gaussian random
numbers, but also by using the sequence of random numbers following:
\begin{eqnarray}
\tilde{W}_i(h) & = & \left \{ \begin{array}{ccc}
                    -\sqrt{3 }, &  & R < 1/6 \\
                    0, &  & 1/6 \le R < 5/6  \\
                    \sqrt{3 }, &  & 5/6 \le R
                    \end{array} \right.
\end{eqnarray}
where $R$ is a uniformly distributed random number on the interval
(0,1). This trick significantly reduces the computational cost in
generating random numbers.

\section{NUMERICAL TESTS}
The above algorithm was tested on a one-dimensional stochastic
harmonic oscillator with a simple form of the multiplicative
noise. The equations of motion were
\begin{eqnarray}
\dot{p} & = & F_1(p,x) + \sigma(x) \xi(t) \nonumber\\
\dot{x} & = & p
\label{testeqn}
\end{eqnarray}
where $F_1(p,x) = -\gamma p -\eta^2 x$ and $\sigma(x)=-\alpha x$. The
stochastic leapfrog integrator for this case is given by
Eqns.~(\ref{walgo}) (white noise)
with the substitutions $x_1=p$, $x_2=x$. 

\begin{figure}
\centering
\includegraphics*[width=55mm,angle=-90]{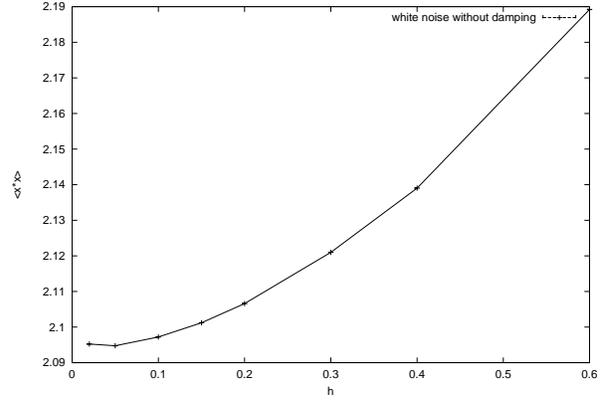}
\caption{Zero damping convergence test. $\langle
x^2(t)\rangle$ at $t=6$ as a function of step size with white
Gaussian noise. Solid lines represent quadratic fits to the data points (diamonds).}
\label{quad}
\end{figure}

\begin{figure}
\centering
\includegraphics*[width=55mm,angle=-90]{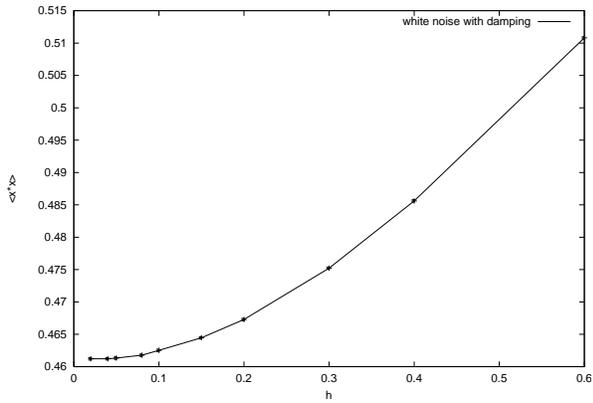}
\caption{Finite damping ($\gamma=0.1$) convergence test. $\langle
x^2(t)\rangle$ at $t=12$ as a function of step size with white
Gaussian noise. Solid lines represent quadratic fits to the data points (diamonds).}
\label{quad_damp}
\end{figure}

\begin{figure}
\centering
\includegraphics*[width=90mm]{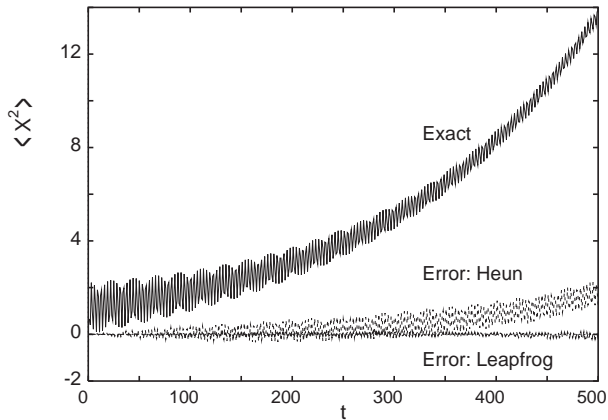}
\caption{Comparing stochastic leap-frog and the Heun algorithm:
$\langle x^2(t)\rangle$ as a function of $t$. Errors are given
relative to the exact solution.}
\label{comp}
\end{figure}

As a first test, we computed $\langle x^2\rangle$ as a function of
time-step size. To begin, we took the case of zero damping constant
($\gamma=0$), where $\langle x^2\rangle$ can be determined
analytically. The curve in Fig.~\ref{quad} shows $\langle
x^2\rangle $ at $t=6.0$ as a function of time-step size with white
Gaussian noise. Here, the parameters $\eta$ and $\alpha$ are set to
$1.0$ and $0.1$. 
The analytically determined value of
$\langle x^2\rangle $ at $t=6.0$ is $2.095222$. 
The quadratic
convergence of the stochastic leap-frog algorithm is clearly seen in
the numerical results.  
We also verified that the quadratic convergence is present for nonzero
damping ($\gamma=0.1$). At $t=12.0$, and with all other parameters as
above, the convergence of $\langle x^2 \rangle$ as a function of time
step is shown by the curve in Fig.~\ref{quad_damp}.
As a comparison against the conventional Heun's algorithm~\cite{greiner}, 
we computed
$\langle x^2 \rangle $ as a function of $t$ using $100,000$ numerical
realizations for a particle starting from $(0.0,1.5)$ in the $(x,p)$
phase space. The results along with the analytical solution and a
numerical solution using Heun's algorithm are given in
Fig.~\ref{comp}. Parameters used were $h=0.1$, $\eta=1.0$, and
$\alpha=0.1$. The advantage in accuracy of the stochastic leap-frog
algorithm over Heun's algorithm is clearly displayed, both in terms of
error amplitude and lack of a systematic drift.
\section{APPLICATION}
\begin{figure}
\centering
\includegraphics*[width=55mm,angle=-90]{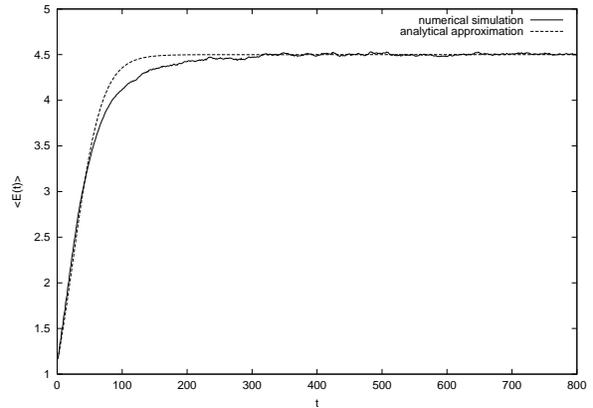}
\caption{Temporal evolution of the scaled average energy $\langle
E(t)\rangle$ with multiplicative noise from numerical simulation
and analytical approximation.}
\label{app1}
\end{figure}

In this section, we apply our algorithm to studying the approach to
thermal equilibrium of an oscillator with multiplicative noise. 
The governing equations are:
\begin{eqnarray}
\dot{p} & = & - \omega_0^2 x - \lambda x^2 p
- \sqrt{2 D} x \xi_2(t) \nonumber\\
\dot{x} & = & p
\label{multle}
\end{eqnarray}
where the diffusion coefficients $D = \lambda k T$,
$\lambda$ is the coupling constant, 
and $\omega_0$ is the oscillator angular
frequency without damping.
In Fig.~\ref{app1}, we display the time evolution of the average
energy with
multiplicative noise from the simulations and the approximate
analytical calculations~\cite{lindenberg}.
The analytic approximation
resulting from the application of the energy-envelope method
is seen to be in reasonable agreement with the numerical
simulations for $kT=4.5$. The slightly higher equilibrium rate from
the analytical calculation is due to the truncation in the energy
envelope equation using the $\langle E^2(t)\rangle\approx 2\langle
E(t)\rangle^2$ relation which yields an upper bound on the rate of
equilibration of the average energy~\cite{lindenberg}. 

\section{Conclusions}
We have presented a stochastic leap-frog algorithm for 
Langevin simulation with multiplicative noise. This method has the
advantages of retaining the symplectic property in the deterministic
limit, ease of implementation, and second-order convergence of moments
for multiplicative noise. Sampling a uniform distribution instead of a
Gaussian distribution helps to significantly reduce the computational
cost.  A comparison with the conventional Heun's algorithm highlights
the gain in accuracy due to the new method. Finally, we have applied
the stochastic leap-frog algorithm to a nonlinear mechanic-oscillator 
system to investigate the 
the nature of the relaxation process.

\section{Acknowledgments}
We acknowledge helpful discussions with Grant Lythe and Robert Ryne.


\begin{thebibliography}{9}
\bibitem{piwinski}A. Piwinski, Proc. 9th Int. Conf. on High Energy
Accelerators, Standord, 1974 (SLAC, Stanford, 1974) p. 405.
\bibitem{risken} H. Risken, {\it The Fokker-Planck Equation: Methods
of Solution and Applications} (Springer, New York, 1989).
\bibitem{jones} M. E. Jones, D. S. Lemons, R. J. Mason, V. A. Thomas,
and D. Winske, J. Comput. Phys. {\bf 123}, 169 (1996). 
\bibitem{zwanzig} R. Zwanzig, J. Stat. Phys. {\bf 9}, 215 (1973).
\bibitem{greiner} A. Greiner, W. Strittmatter, and J. Honerkamp,
J. Stat. Phys. {\bf 51}, 94 (1988).
\bibitem{lindenberg} K. Lindenberg and V. Seshadri, Physica {\bf 109}
A, 483 (1981).
\end{thebibliography}
\end{document}